\documentclass[twocolumn,epjc3,dvipsnames]{svjour3}

\usepackage{amsmath}
\usepackage{amssymb}
\usepackage{booktabs}
\usepackage{braket}
\usepackage{cancel}
\usepackage{dcolumn}
\usepackage{graphicx}
\usepackage[utf8]{inputenc}
\usepackage{siunitx}
\usepackage{xcolor}
\usepackage{xspace}
\usepackage{hyperref}
\usepackage{cuted}

\newcommand{\GeV}{\ensuremath{\,\mathrm{GeV}\xspace}}
\newcommand{\MeV}{\ensuremath{\,\mathrm{MeV}\xspace}}

\DeclareMathOperator{\atanh}{atanh}

\renewcommand{\Im}{\operatorname{Im}}

\newcommand{\ie}{\textit{i.e.}\xspace}

\newcommand{\eetohadrons}{\ensuremath{e^+e^- \to \text{hadrons}}\xspace}
\newcommand{\eetoopencharm}{\ensuremath{e^+e^- \to \text{open charm}}\xspace}
\newcommand{\nonDDbar}{\ensuremath{\textrm{non-}D\bar{D}}\xspace}

\newcommand{\Kmatrix}{$K$-matrix\xspace}
\newcommand{\Smatrix}{$S$-matrix\xspace}
\newcommand{\RiemannI}{\ensuremath{\textrm{I}}\xspace}
\newcommand{\RiemannII}{\ensuremath{\textrm{II}}\xspace}

\newcommand{\dynesty}{\texttt{dynesty}\xspace}
\newcommand{\EOS}{\texttt{EOS}\xspace}

\makeatletter
\let\oldtheequation\theequation
\renewcommand\tagform@[1]{\maketag@@@{\ignorespaces#1\unskip\@@italiccorr}}
\renewcommand\theequation{(\oldtheequation)}
\makeatother

\begin{document}

\title{Analysis of the $\boldsymbol{\psi(3770)}$ resonance in line with unitarity and analyticity constraints}
\author{%
    Christoph Hanhart\thanksref{aff:FZJ} \and
    Stephan K\"urten\thanksref{aff:TUM,aff:HISKP} \and
    M\'eril Reboud\thanksref{aff:Siegen} \and
    Danny van Dyk\thanksref{aff:IPPP}
}
\institute{%
\setlength{\parindent}{0pt}%
\label{aff:FZJ}%
Institute for Advanced Simulation and Institut für Kernphysik, Forschungszentrum Jülich, D-52425 Jülich, Germany\and
\label{aff:TUM}%
Physik Department T31, Technische Universit\"at M\"unchen, D-85748 Garching, Germany\and
\label{aff:HISKP}%
Helmholtz-Institut für Strahlen- und Kernphysik (Theorie) and\\ Bethe Center for Theoretical Physics, Universität Bonn, D-53115 Bonn, Germany\and
\label{aff:Siegen}%
Theoretische Physik 1, Naturwissenschaftlich-Technische Fakult\"at, Universit\"at Siegen, D-57068 Siegen, Germany\and
\label{aff:IPPP}%
Institute for Particle Physics Phenomenology and Department of Physics, Durham University, Durham DH1 3LE, UK
}

\dedication{EOS-2023-02,IPPP/23/67,P3H-23-091,SI-HEP-2023-26,TUM-HEP-1480/23}

\maketitle

\begin{abstract}
We study the inclusive and exclusive cross sections of $e^+e^-\to \text{hadrons}$
for center-of-mass energies between $3.70\,\GeV$ and $3.83\,\GeV$ to infer the mass, width, and
couplings of the $\psi(3770)$ resonance.
By using a coupled-channel $K$-matrix approach, we setup our analysis to
respect unitarity and the analyticity properties of the underlying scattering amplitudes.
We fit several models to the full dataset and identify our nominal results through a statistical model comparison.
We find that, accounting for the interplay between the $\psi(2S)$ and the $\psi(3770)$, 
no further pole is required to describe the $\psi(3770)$ line shape.
In particular we derive from the pole location $M_{\psi(3770)}~=~3778.8~\pm~0.3~\MeV$
and $\Gamma_{\psi(3770)}~=~25.0~\pm~0.5~\MeV$.
Moreover, we find the decay to $D^+D^-$ and $D^0\bar{D}^0$ to be consistent with isospin symmetry and
derive an upper bound on the branching ratio $\mathcal{B}(\psi(3770) \to \textrm{non-}D\bar{D}) < 6\%$ at $90\%$ probability.
\end{abstract}

\section{Introduction}

The study of \eetohadrons processes has been useful to improve our understanding of a variety of aspects of particle physics in general and
the strong interaction in particular. These include the confirmation of three as the 
number of strong charges (colours)~\cite{Bar:2001qk},
the
discovery of exotic states
outside the established quark model (see Refs.~\cite{Lebed:2016hpi,Esposito:2016noz,Olsen:2017bmm,Guo:2017jvc,Brambilla:2019esw,Chen:2022asf}
for recent reviews),
and the data-driven prediction of hadronic contributions to the anomalous magnetic moment of the muon~\cite{Aoyama:2020ynm}.

In this analysis, we study \eetoopencharm processes in the immediate vicinity of the $D^0\bar{D}^0$
and the $D^+D^-$ thresholds but below the $D\bar{D}^* + \text{h.c.}$ threshold.
Our study of \eetoopencharm data is motivated by the following questions:
\begin{enumerate}
    \item What is the nature of the $\psi(3770)$ state? To that end, does it decay sizeably into \nonDDbar final states, in contradiction with being a pure $c\bar{c}$ quarkonium state and in support of alternative interpretations?
    \item Are contemporary theoretical frameworks capable to describe the
    now-available high-resolution measurements of $\eetoopencharm$ processes?
    \item How many vector states are necessary to descibe the data within the mass range studied? 
    \item \label{item:intro:model} Can we describe the \eetoopencharm spectrum well enough to use it for data-driven
    predictions of non-local contributions in $b\to s\ell^+\ell^-$ processes?
\end{enumerate}
A previous study covering a large part of the \eetoopencharm phase space has been carried out in Ref.~\cite{Shamov:2016mxe}; it uses,
amongst others, high resolution BES, BESII, and BESIII data.
That study uses a model consisting of a sum of Breit-Wigner functions.
This approach is known to violate unitarity of the \Smatrix in the description of broad resonances close to their dominant decay threshold (see Review {\it Resonances} in Ref.~\cite{Workman:2022ynf}), which clearly holds for the $\psi(3770)$.
As a consequence, the line shape extracted from \eetoopencharm data cannot be transferred
to other applications, such as data-driven predictions of $b\to s\ell^+\ell^-$ decays, without incurring an unquantifiable model uncertainty.
To overcome this issue, we strive to model the relevant scattering amplitudes with as few assumptions as possible before fitting our models to the available data.
Our choice of phase space window implies the absence of dominant three-hadron final states.
This is a necessary prerequisite for the \Kmatrix framework, which we use in this study.
A previous \Kmatrix analysis of exclusive \eetoopencharm data has been carried out in Ref.~\cite{Uglov:2016orr},
exclusively using Belle data.
This data covers a much larger energy range than what we study here but features a substantially lower resolution than the BES data.
It is therefore interesting to see if the available high-resolution measurements by the BES, BESII, and BESIII experiments can be described within the highly-predictive \Kmatrix framework.
Moreover, we allow for the $\psi(2S)$ to interfere with the $\psi(3770)$, which appears necessary
to describe the data.

Conceptually our work seems similar to that of Ref. \cite{Coito:2017ppc},
however, we deviate in a couple of crucial points: we allow for \nonDDbar decays and for a contribution
of the $\psi(2S)$.
The most salient difference is that our framework does not generate additional poles beyond those explicitly included by construction.
A more detailed comparison to the results of  Ref.~\cite{Coito:2017ppc}
will be presented below.

The structure of this article is as follows. We discuss our analysis setup in \autoref{sec:setup},
including a brief overview of the \Kmatrix framework, a description of the
available experimental data, and the definition of our fit models.
We present the numerical results in \autoref{sec:results}.
A summary and outlook follows in \autoref{sec:summary}.
We describe a path toward data-driven predictions of the non-local
form factors in rare semileptonic $b$ decays in \autoref{app:non-local}.

\section{Setup}
\label{sec:setup}

\subsection{Analysis Framework}
\label{sec:setup:framework}
The \Kmatrix framework has first been proposed in Ref.~\cite{Chung:1995dx}
to describe $2 \to r \to 2$ scattering amplitudes and $r\to 2$ decay amplitudes, where $r$ denotes some hadronic resonance.
The framework allows straightforwardly for the inclusion of two-body channels
and automatically leads to unitary amplitudes.
Here, we apply the \Kmatrix framework in its modern, Lorentz-invariant
form; see Ref.~\cite{Workman:2022ynf} for a review and a collection of the relevant formulae.\\

In the \Kmatrix framework, a scattering amplitude $\mathcal{M}$ is modelled as
\begin{equation}
\label{eq:def:scattering_amplitude}
    \mathcal{M} = n\left[ 1 - \mathcal{K} \, \Sigma \right]^{-1} \mathcal{K} \, n.
\end{equation}
Here, columns and rows of $\mathcal{M}$ correspond to the initial and final states
of the processes under consideration, which are commonly referred to as ``channels''.
The same holds for the columns and rows of the underlying matrix $\mathcal{K}$.
Moreover, to ensure unitarity of the \Smatrix and to uphold symmetry under time-reversal,
$\mathcal{K}$ must be real-valued and symmetric, respectively.
The channels' vertex structure is accounted for by the diagonal matrix
$n=\mathrm{diag}(n_a, n_b, ...)$ with
\begin{align}
    \label{eq:def:n}
    n_k = (q_k / q_0)^{l_k} F_{l_k} (q_k / q_0)\,.
\end{align}
In the above, $l_k$ is the orbital angular momentum in channel $k$ and 
\begin{equation}
    q_k(s) = \frac{\lambda(s, M_{k1}^2, M_{k2}^2)^{1/2}}{2\sqrt{s}}
\end{equation}
is the break-up momentum, expressed in terms of the Källén triangle function.
The masses of the two hadrons of channel $k$ are denoted by $M_{k1}$ and $M_{k2}$, respectively.
Their break-up momentum is further used to define a channel's phase space function
$\rho_k = q_k(s) / (8 \pi \sqrt{s})$\,.
Moreover, $q_0$ is some fixed momentum scale, conventionally chosen between $0.2\GeV$ and $1\GeV$~\cite{Workman:2022ynf,Achasov:2021adv},
and $F_{l_k}$ are the Blatt-Weisskopf form factors~\cite{Blatt:1952ije}
\begin{equation*}
     F_0^2(z) = 1 \,, \qquad F_1^2(z) = 1 / (1 + z^2)\,.
\end{equation*}
The matrix $\Sigma$ in \autoref{eq:def:scattering_amplitude} is a diagonal matrix
$\Sigma = \mathrm{diag}(\Sigma_a(s),\Sigma_b(s),...)$,
where the functions $\Sigma_k(s)$ are channel-specific, modified Chew-Mandelstam functions.
The latter functions are the proper analytic completions of the phase space factors
$i \rho_k(s) n_k(s)^2$ by means of dispersion integrals,
which allow for the
continuations of the amplitudes into the complex plane.
Here, we are only concerned with channels for which $M_{k1} = M_{k2}$, which is reflected in the
formulas for the modified Chew-Mandelstam functions.
For an $S$-wave channel (i.e., $l_k = 0$), they read
\begin{equation}
    \Sigma_{k}(s) = \frac{1}{8 \pi^2} \Pi_0 \,.
\end{equation}
For a $P$-wave channel (i.e., $l_k = 1$) they read
\begin{equation}
    \Sigma_{k}(s) = \frac{1}{8\pi^2} \frac{s - s_\mathrm{th}}{s_0} \left( F_1^2(q_k(s)/q_0) \, \Pi_0(s) + \Pi_1(s) \right).
\end{equation}
In the above we use
\begin{align}
    \Pi_0 & = - \frac{\sqrt{s_\mathrm{th} - s}}{\sqrt{s}} \, \arctan \sqrt{\frac{s}{s_\mathrm{th} - s}}, \\
    \Pi_1 & = \frac{s_0^{3/2}}{\sqrt{s_0 - s_\mathrm{th}} (s + s_0 - s_\mathrm{th})} \, \atanh \sqrt{1 - \frac{s_\mathrm{th}}{s_0}},
    \label{eq:def:Pi1}
\end{align}
where $s_\mathrm{th} = 4\, M_k^2$ and $s_0 = 4\, q_0^2$.
Note that the pole in \autoref{eq:def:Pi1} cancels exactly the pole due to $F_1^2$,
which makes both $\Sigma_{k}$ analytic functions of $s$ in the whole complex plane,
except for a branch cut starting at $s=s_\mathrm{th}$.
This branch cut connects the two Riemann sheets of the Chew-Mandelstam functions. The formulas above are suitable to evaluate them on their first Riemann sheet only.
To evaluate the function on their second Riemann sheet, we use
\begin{equation}
    \Sigma^{\RiemannII}_{k}(s) = \Sigma_{k}(s) + 2 i \left( \rho_k(s^*) \, n_k^2(s^*) \right)^*.
\end{equation}

Following Ref.~\cite{Workman:2022ynf}, we parametrize the \Kmatrix as follows:
\begin{equation}
\label{eq:def:K-param}
    \mathcal{K}_{ij}(s)
        = \sum_{r=1}^{N_R}\frac{g_i^r g_j^r}{m_r^2-s}
        + c_{ij} \, .
\end{equation}
The first term describes the $N_R$ resonances included explicitly in the model,
with bare mass $m_r$ and 
$g_i^r$ for their coupling to the channel $i$, all of them real valued.
The second term is the background constant that models non-resonant contributions
of, e.g., tails
of resonances outside the phase space window considered here.
\\

Each resonance $r$ gives rise to pairs of poles of the scattering amplitudes
\autoref{eq:def:scattering_amplitude} on the unphysical Riemann sheets.
For $N_C$ channels, this amounts to a total of $2^{N_C}$ Riemann sheets.
However, given the parametrisation employed here,
it is sufficient to continue the individual self-energies $\Sigma_k$ to
their second sheet to reach those poles. We may label any given sheet with a multi index, by denoting on which
sheet the respective self-energy for each channel is evaluated.
In this notation, the
physical sheet is denoted as $\vec{\RiemannI} \equiv \{\RiemannI,\dots, \RiemannI\}$.
The resonance pole located closest to the physical axis is commonly quoted as the resonance
pole and parametrised as
\begin{equation}
    \sqrt{s_r} = M_r - i \frac{\Gamma_r}{2} \,,
\end{equation}
which defines the resonance's physical mass $M_r$ and total decay width $\Gamma_r$.
To access these properties, one requires the numerical evaluation of the scattering
amplitudes on the proper Riemann sheet.
In our analysis, we are interested only in the description of the $\psi(3770)$ pole,
which is located above all modelled hadronic thresholds.
To determine this pole's properties, it therefore suffices to consider the
Riemann sheet closest to the physical axis, which we denote as
$\vec{\RiemannII} \equiv \{ \RiemannII, \dots, \RiemannII \}$.

This sheet can be reached by means of
\begin{equation}
    \mathcal{M}^\RiemannII = n\left[ 1 - \mathcal{K} \, \Sigma^{\vec{\RiemannII}} \right]^{-1} \mathcal{K} \, n\, ,
\end{equation}
where $\Sigma^{\vec{\RiemannII}}$
denotes the self-energy matrix with
all channel-self-energies continued to their second sheet.
To determine the physical quantities, such as partial decay widths and branching ratios,
we require access to the renormalized couplings $G_k^r$.
We extract these couplings as residues of the diagonal elements in channel space
of a partial-wave amplitude on the
proper Riemann sheet
\begin{equation}
    (G_k^r)^2 = -\frac{1}{2\pi i} \oint_{C(s_r)} ds \, \mathcal{M}^{\RiemannII}_{kk}(s).
\end{equation}
Here $C(s_r)$ describes a contour around the resonance's pole position, $s_r$, on the proper Riemann sheet
 that avoids all other singularities.
The definition of the physical observables then reads
\begin{equation}
    \Gamma_{r\to a} = \frac{|G_a^r|^2}{M_r} \rho_a(M_r^2) \quad \text{and} \quad \mathcal{B}_{r \to a} = \frac{\Gamma_{r\to a}}{\Gamma_r} \ ,
\end{equation}
where we employed the narrow width approximation for the calculation of
the partial width.
Note that we do not impose the identity $\Gamma_r = \sum_a \Gamma_{r\to a}$.
We discuss this type of relation
later on in \autoref{sec:setup:analysis}.
Finally, we compute the cross sections from the scattering amplitudes as 
\begin{equation}
    \sigma_{e^+e^- \to k}(s) = \frac{1}{16 \pi s} \, \frac{\rho_k(s)}{\rho_{e^+e^-}(s)} \, \frac{\mathcal{N}_k}{4} \, \left| \mathcal{M}_{e^+e^-,k} \right|^2,
\end{equation}
where $\mathcal{N}_k = 2 l_k + 1$ is a combinatorial factor and the factor of 4 accounts for the number of
spin configurations in the initial state.\\

\paragraph{Resonances}
For this analysis, we study cross sections for exclusive $\eetoopencharm$ processes.
All resonances must share the same quantum numbers as the photon, i.e., all flavour
quantum numbers must vanish
and $J^{PC}=1^{--}$, where $J$ denotes the total angular momentum.
The energy range of interest here, $4 M_{D^0}^2 < s < (M_D + M_{D^*})^2$,
sits above the well-known narrow charmonium resonances $J/\psi$ and $\psi(2S)$
and is dominated by effects of the broad $\psi(3770)$ resonance.
We do not aim at modelling the shape of the $J/\psi$ and $\psi(2S)$ resonances.
Nevertheless, the interference effect between the $\psi(2S)$ and the $\psi(3770)$
is found to play a major role in the shape of the $\psi(3770)$
in various works~\cite{Uglov:2016orr,Shamov:2016mxe}.
Hence, we include the $\psi(2S)$ as the closest narrow charmonium state in our model:
\begin{align}
\label{eq:def:resonances}
    r \in \lbrace \psi(2S), \psi(3770)\rbrace \quad \text{and} \quad N_R=2 \ .
\end{align}

\paragraph{Channels}
The energy range of interest overlaps with only a small slice of the full phase space of open-charm production.
The dominant processes are therefore $e^+e^-$ $\to \nonDDbar$, $e^+e^- \to D^+D^-$, and $e^+e^- \to D^0\bar{D}^0$.
A comment is due on the hadronic $\nonDDbar$ final states.
Empirically, it is known that various genuine non-two-body final states contribute here~\cite{Workman:2022ynf}
that cannot be straightforwardly expressed within the \Kmatrix framework as applied here~\cite{Chung:1995dx,Workman:2022ynf}.
For our purpose, this inclusive final state is expected to yield a numerically dominant contribution only to the decay width of the $\psi(2S)$ resonance, \ie, well below the open charm threshold.
We therefore setup our model using the following assumptions:
\begin{itemize}
    \item The effects of the $\psi(2S)$ modify the line shape of the $\psi(3770)$ and a description of this
    modification is needed. However, we are not interested in describing the line shape of the $\psi(2S)$.
    For the purpose of determining the impact on the $\psi(3770)$ line shape through interference,
    we model this component as an effective $P$-wave two-body channel $\text{eff}_{\psi(2S)}$ with threshold $4 M_\pi^2$.
    Note that the results are insensitive to the concrete value chosen here as long as it is located significantly below the energy range considered.\\
    Moreover, we allow for a non-vanishing $\nonDDbar$ component to the decay width of the $\psi(3770)$.
    For the purpose
    of determining the overall width of the $\psi(3770)$ we model this component as an effective $P$-wave two-body
    channel $\text{eff}_{\psi(3770)}$ with threshold $4 M_\pi^2$.
    We study two scenarios: one in which $\text{eff}_{\psi(3770)}$ and $\text{eff}_{\psi(2S)}$ are assumed
    to be distinct and hence non-interfering; and one in which the channels are identical,
    $\text{eff}_{\psi(3770)}$ $= \text{eff}_{\psi(2S)} = \text{eff}_{\psi}$.
    \item The cross sections in our phase space windows are dominated by $D^+D^-$ and $D^0\bar{D}^0$ final states.
    We model these final states via two independent $P$-wave channels (i.e., $l_{D^+D^-} = l_{D^0\bar{D}^0} = 1$).
    \item The coupling of the two resonances to $e^+e^-$ enter all cross sections discussed here. To keep our
    numerical code as simple as possible, we define a \Kmatrix channel with label $e^+e^-$. This approach leads
    to an inadvertent accounting for hadronic open-charm contributions to the $e^+e^-$ vacuum polarisation,
    which is negligible in our case. We have checked that our numerical code yields virtually indistinguishable
    results compared to a (simpler) code that uses a $P$-vector approach for the $e^+e^-$ channel.
    We model the $e^+e^-$ initial state as an $S$-wave channel (i.e., $l_{e^+e^-} = 0$).
\end{itemize}
This leaves us with the following sets of channels,
depending on the number of non-$D\bar D$ channels
included. Each channel features an independent set of couplings. We thus have either $N_C = 5$ with
\begin{align}
\label{eq:def:channels}
    k \in \lbrace e^+e^-, D^+D^-, D^0\bar{D}^0, \text{eff}_{\psi(2S)}, \text{eff}_{\psi(3770)} \rbrace\,,
\end{align}
or $N_C = 4$ with
\begin{align}
\label{eq:def:channels2}
    k \in \lbrace e^+e^-, D^+D^-, D^0\bar{D}^0, \text{eff}_{\psi} \rbrace\,.
\end{align}

\subsection{Experimental Data}
\label{sec:setup:data}

Experimental measurements of the $e^+e^-\to \textrm{hadron}$ cross sections in the energy
range of interest
are available from the BaBar~\cite{Aubert:2008pa},
Belle~\cite{Pakhlova:2008zza}, BES~\cite{Bai:2001ct}, BESII~\cite{Ablikim:2006mb}, BESIII~\cite{BESIII:2021wib}, and CLEO~\cite{CroninHennessy:2008yi} experiments.
These measurements vary strongly in the underlying approaches to measure the cross sections, which can
roughly be divided into two categories:
\begin{description}
    \item[\textit{energy scan}] The BES, BESII, BESIII, and CLEO experiments take data at a variety of different
    center-of-mass energies, $\sqrt{s}$, of the $e^+e^-$ collisions. This enables them to obtain measurements
    of the exclusive cross sections at different values of $\sqrt{s}$.
    The resolution of these data points is $\lesssim 10\,\MeV$, yielding high-resolution
    measurements of the spectra.
    In the context of this analysis, we treat energy-scan measurements as single-points with vanishing bin width.
    
    \item[\textit{initial-state radiation}] The BaBar and Belle experiments work at fixed center-of-mass
    energies, $\sqrt{s}~\sim~10\GeV$, far above the energy range of interest.
    Nevertheless, they can access lower energies
    by means of initial-state radiation (ISR), i.e., radiation of an energetic photon off either of the
    initial-state leptons. This approach does not permit a high-resolution energy scan of the
    pertinent cross section. Instead, those results are presented as integrated cross sections
    in relatively coarse bins of the center-of-mass energy.
\end{description}
For this analysis, we use only the measurements by the BES, BESII, and BESIII experiments. Our reasoning is
as follows:
\begin{itemize}
    \item The BES, BESII, and BESIII measurements are based on much larger data sets than the CLEO measurements.
    Consequently, the latter are not competitive with the former within our analysis on account of larger statistical uncertainties.
    \item The BES, BESII, and BESIII measurements provide a high-resolution access to the
    energy dependence of the exclusive cross sections. The BaBar and Belle results cannot compete
    with these BES results due the limitations of the ISR method.
\end{itemize}
We refer to the data sets on the ratio $R= \sigma(e^+e^- \to \textrm{hadrons})/\sigma(e^+e^- \to \mu^+\mu^-)$ as inclusive data and to the data sets on $e^+e^- \to D^0 \bar{D}^0$ and $e^+e^- \to D^+ D^-$ as the exclusive data.
Taking the exclusive data into account allows our fit to be sensitive to isospin symmetry violation.
We only use data points with center-of-mass energy $\sqrt{s} \leq 3.83\GeV$, to limit the experimental pollution of the $\psi(4040)$ resonance.
This leaves us with the following combined dataset that is used throughout our analyses:
\begin{description}
    \item[\textit{inclusive}] We use $12$ and $60+1$ experimental measurements from the analyses by BES~\cite{Bai:2001ct} and BESII~\cite{Ablikim:2006mb,BES:2006dso}, denoted as \texttt{BES\ 2002}, \texttt{BESII\ 2006A} and \texttt{BESII\ 2006B}, respectively, in the rest of this paper;
    \item[\textit{exclusive}] We use $26$ and $27$ experimental measurements from a preliminary BESIII analysis~\cite{Julin:2017jcl}
    that we will denote as \texttt{BESIII\ 2017} in the following.
    We do not account for small systematic correlations between the $D^+D^-$ and $D^0\bar{D}^0$ final states.
    The observed cross section $\sigma^\text{obs}$ still needs to be converted to the Born
    cross section $\sigma^\text{B}$. This is achieved by~\cite{Husken:2024hmi}
    \begin{equation*}
        \sigma^\text{B}(E) = \sigma^\text{obs}(E) \frac{|1 - \Pi(E)|^2}{1 + \delta(E)},
    \end{equation*}
    where $\Pi(E)$ is the vacuum polarization and $\delta(E)$ is the radiative correction
    that accounts for initial-state radiation.
    This is done to ensure consistency of our analysis with
    respect to the inclusive cross section measurements.
\end{description}
This corresponds to a total of $126$ observations.
As they are measured during different experimental runs, all these measurements are statistically independent.
The systematic uncertainties are provided in the experimental publication.
They permit us to reconstruct the full correlation matrices by separating the energy-independent uncertainties from the other systematic uncertainties.\\

We fix the value of the $R$ ratio below the open-charm threshold to the value $R_{uds} = 2.171$~\cite{Harlander:2002ur}.
To ensure the convergence of the fits and the physical meaning of the models, we furthermore consider two additional constraints:
\begin{itemize}
    \item The bare partial width of the $\psi(2S)$ resonance to $e^+e^-$ is constrained to $\Gamma_{\psi(2S)\to e^+e^-} = (2.33 \pm 0.04)$ keV.
    This constraint has a limited impact on the fit and is just used to ensure convergence.

    \item The value of the $R$ ratio far above the open-charm threshold should not exceed the value $R_{udsc} = 3.55$
    \cite{Harlander:2002ur}. To implement this constraint, in the fit
    we impose a penalty function
    \begin{equation}
        -2\log P \supseteq \frac{(r - 3.55)^2}{\sigma^2} \theta(r - 3.55),
    \end{equation}
    where $r = R(\sqrt{s} = 9\GeV)$ corresponds to the four-flavour $R$ ratio evaluated below the first $b\bar{b}$ resonance
    and $\theta$ is the Heaviside function.
    We use $\sigma = 10\%$ to account for the theory uncertainty of the $R$ ratio prediction.
    Here again, the fit is not sensitive to these exact values, but using this prior ensures that the model remains physical. 
\end{itemize}

\subsection{Analysis}
\label{sec:setup:analysis}

To confront our physical model with the available data, we perform a Bayesian analysis.
Central to this type of analysis is the posterior probability density function (PDF)
of our fit parameters $\vartheta$,
\begin{equation}
    P(\vartheta \,|\, D, M) = \frac{P(D, M \,|\, \vartheta)\, P_0(\vartheta \,|\, M)}{Z(D, M)}\,.
\end{equation}
In the above, $P(D, M \,|\, \vartheta)$ is known as the (experimental) likelihood,
$P_0$ is the prior PDF of our parameters, and the evidence $Z(D, M)$
ensures the normalization of the posterior PDF. The label $D$ refers to the
dataset used in the fit (see \autoref{sec:setup:data}) and the label $M$ refers to the fit model (discussed below).

Our fit parameters can be classified as follows:
\begin{description}
    \item[\textit{masses}] We fix the bare mass of the $\psi(2S)$ to the physical world average $M_{\psi(2S)} = 3.6861 \GeV$~\cite{Workman:2022ynf}.
    We fit the bare mass parameter of the $\psi(3770)$. This amounts to one fit parameter.
    \item[\textit{couplings}] We fit the bare couplings of all resonances $r$ listed in \autoref{eq:def:resonances}
    to the channels listed in \autoref{eq:def:channels} or \autoref{eq:def:channels2},
    depending on the fit model. In the former setting
    the $\psi(2S)$ does not couple to the channel $\text{eff}_{\psi(3770)}$ and vice versa.
    In the latter both vector resonances couple to the same channel. In both cases
     this amounts to eight parameters describing the bare couplings.
    \item[\textit{background terms}] We fit the background terms introduced in \autoref{eq:def:K-param}.
    In our analysis, only background terms for the processes $e^+e^- \to \lbrace D^0\bar{D}^0, D^+D^-\rbrace$
    are considered. Symmetry of the \Kmatrix implies that we must use the same background terms
    for the time-reversed processes.
    This amounts to two independent fit parameters.
    \item[\textit{effective momentum}] We fit the effective momentum $q_0$ entering \autoref{eq:def:n}.
    Although this quantity is a-priori channel dependent, we use a common value for $q_0$ across all
    channels. This amounts to one fit parameter.
\end{description}
By construction, all fit parameters are real-valued parameters as demanded
by the properties of $\mathcal{K}$; see \autoref{sec:setup:framework}.
We find that the likelihood (and hence the posterior PDF) exhibits several symmetries with
respect to the above parameters that help in reducing the prior ranges of our analysis:
\begin{itemize}
    \item If the effective channels are specific to a single resonance only and we do not impose a background term for them,
    the posterior PDF is insensitive to the signs of the effective couplings.
    In that case, we can choose both couplings to be positive.
    If, on the other hand, the effective channels are allowed to interfere, the relative sign
    between both couplings becomes observable.
    Hence, we choose the coupling to the $\psi(3770)$ to be positive.
    \item The posterior PDF is insensitive to the overall sign of the full set of bare couplings to a common resonance $r$, since each observable contains the product of two resonance couplings.
    Put differently, we can change the sign of all bare couplings $g_k^r$ for a fixed $r$
    without changes to the posterior PDF. This enables us to choose the sign of one bare coupling
    per (fixed) resonance. We choose the couplings $g_{e^+e^-}^r$ to be positive.
    \item The posterior PDF is insensitive to the overall sign of the full set of couplings to a common single channel $k$. Put differently, we can change the sign of all bare couplings $g_k^r$ for a fixed $k$
    without changes to the posterior PDF. This enables us to choose the sign of one bare coupling
    per (fixed) channel $k$. We choose the coupling $g_k^{\psi(3770)}$ to be positive.
\end{itemize}
We use as the prior PDF a product of uniform PDFs for each fit parameter.\\

We define the following fit models that are investigated as part of our analysis:
\begin{description}
    \item[\textit{minimal}] We fit the $\psi(3770)$ bare mass parameter and seven bare coupling parameters
    for the channels discussed above, fixing the coupling of the \nonDDbar component of the $\psi(3770)$ (modelled by the $\text{eff}_{\psi(3770)}$ channel) to zero.
    (8 parameters)
    \item[\textit{no background}] Same as the ``minimal'' model.
    We additionally fit the effective $\text{eff}_{\psi(3770)}$ channel. (9 parameters)
    \item[\textit{background}] Same as the ``no background'' model.
    We additionally fit the constant background parameter in the off-diagonal \Kmatrix entries for the $e^+e^- \to D^0\bar{D}^0$ and $e^+e^-\to D^+D^-$ processes.
    Since our framework is constructed to produce a symmetric \Kmatrix,
    these background terms also contribute to the time-reversed processes $D^0\bar{D}^0 \to e^+e^-$ and $D^+D^-\to e^+e^-$. (11 parameters)
    \item[\textit{$q_0$ variation}] Same as the ``background'' model.
    We additionally fit the effective scale $q_0$, assuming, as stated above, that this parameter is the same for all the channels. (12 parameters)
    \item[\textit{interference}] We fit the $\psi(3770)$ bare mass parameter and the eight bare coupling parameters as discussed above
    in the context of one joint effective channel with couplings to both the $\psi(2S)$ and the $\psi(3770)$,
    see \autoref{eq:def:channels2}. (11 parameters)
\end{description}

To carry out our analysis we use the \EOS software~\cite{EOSAuthors:2021xpv} in version 1.0.11~\cite{EOS:v1.0.11}, which has been modified for this purpose.
Our analysis involves the optimisation of the posterior to determine the best-fit point or points.
Since all experimental measurements used here are represented by a Gaussian likelihood, we compute
the global $\chi^2$ value in the best-fit point(s), providing a suitable test statistic for the fit.

We further produce importance samples of the model parameters for each fit model. This enables
us to produce posterior-predictive distributions for dependent observables, including those used
in the likelihood but also observables that are as-of-yet unmeasured. We produce the importance
samples by application of the dynamical nested sampling algorithm~\cite{Higson:2018}.
To this end, \EOS interfaces with the \dynesty software~\cite{Speagle:2020,dynesty:v2.0.3}.
Usage of dynamical nested sampling provides the additional benefit of estimating the evidence $Z(D, M)$
in parallel to sampling from the posterior density. This enables us to carry out a Bayesian
model comparison between two models $M_1$ and $M_2$ for a common dataset $D$
through computation of the Bayes factor
\begin{equation}
    B(M_2, M_1) \equiv \frac{Z(D \,|\, M_2)}{Z(D \,|\, M_1)}\,.
\end{equation}
A Bayes factor larger than unity favours model $M_2$ over model $M_1$.
Jeffreys provides a more detailed interpretation of the Bayes factor~\cite{Jeffreys:1939xee}.

\paragraph{Pole position} To determine the position of the $\psi(3770)$ pole in the complex plane,
we carry out a root finding procedure for $\det \left[1 - \mathcal{K} \, \Sigma^{\vec \RiemannII}\right]$.
To determine the uncertainty on the pole position, we repeat the procedure for each posterior sample.

\paragraph{Viability tests} To test the accuracy of our numerical implementation, we perform three types of viability tests a-posteriori.
\begin{itemize}
    \item Since our setup respects the unitarity of the \Smatrix, we expect the sum of the partial decay widths
to correspond to the total decay width, within the uncertainties of the fit.
    \item Since final state interaction is a long-distance effect, we expect the short-distance dominated
    residues of the resonance poles to factorize:
    \begin{equation}
        -\frac{1}{2\pi i} \oint_{C(s_r)} \mathcal{M}^\RiemannII_{a b}(s) \, ds
            = G_a^r \times G_b^r\,.
    \end{equation}
    We remind that we extract the physical couplings $G_k^r$ from their respective partial wave amplitudes
    $\mathcal{M}^\RiemannII_{kk}(s)$.
    \item The spectral function of the $\psi(3770)$ defined as~\cite[chapter 10.7]{Weinberg:1995mt}
        \begin{eqnarray}\nonumber
            \text{spect}_{\psi(3770)}(s) &=& - \frac{1}{\pi}\\ & & \hspace{-2.5cm}
            \times
            \Im\Bigg[\frac{1}{s {-} m^2_{\psi(3770)} {+} \sum
\left(
g_k^{\psi(3770)}
\right)^2 
 \Sigma_k(s)}\Bigg] ,
        \end{eqnarray}
    must be normalised, (i.e.) it must fulfill the property
    \begin{equation}
        \int_{s_\mathrm{th}}^\infty \text{spect}_{\psi(3770)}(s) \, ds = 1 \,,
    \end{equation}
    where $s_\mathrm{th}$ is the first hadronic threshold.
    
\end{itemize}
Significant violation of either test would indicate potential issues with the numerical implementation of our framework.
We apply these tests a-posteriori only, since the information needed to perform the test is not readily accessible in the course of the optimization of or the sampling from the posterior density.
A numerical implementation may violate these tests due to loss of precision or use of functions outside their domain.
This is meant as a practical test of the implementation, not a test of the physics.

\section{Results and Interpretation}
\label{sec:results}

\begin{table*}[t]
    \centering
    \resizebox{\textwidth}{!}{
    \begin{tabular}{lccccccc}
        \toprule
        Model           & $\chi^2$ & d.o.f. & $p$-value $[\%]$ & $\log(Z)$ & $M_{\psi(3770)}$ [MeV]
        & $\Gamma_{\psi(3770)}$ [MeV] & $\mathcal{B}_{\nonDDbar}$ [\%]\\
        \midrule
        minimal         & 120 & 119 & 46.0 & 82.0 & $3779.0 \pm 0.3$ & $23.5 \pm 0.4$ & --- \\
        no background   & 120 & 118 & 44.0 & 79.0 & $3778.9 \pm 0.3$ & $23.6 \pm 0.4$ & $<6.1$ \\
        background      & 107 & 116 & 71.8 & 81.7 & $3778.8 \pm 0.3$ & $25.0 \pm 0.5$ & $<5.8$ \\
        $q_0$ variation & 106 & 115 & 71.8 & 69.3 & $3778.8 \pm 0.3$ & $24.6 \pm 0.6$ & $<5.0$ \\
        interference    & 107 & 116 & 71.5 & 80.5 & $3778.8 \pm 0.3$ & $25.0 \pm 0.5$ & $<6.1$ \\
        \bottomrule
    \end{tabular}
    }
    \caption{\label{tab:results:gof}
        Summary of the analysis of each model.
        d.o.f. refers to the degrees of freedom, $\log(Z)$ to the Bayesian (natural-)log-evidence
        and $\mathcal{B}_{\nonDDbar}$ stands for $\mathcal{B}(\psi(3770)\to\nonDDbar)$.
        In the last column, upper bounds are given at 90\% probability.
        }
\end{table*}

We perform a total of five analyses, using the dataset described in \autoref{sec:setup:data} and the five fit models
described in \autoref{sec:setup:analysis}. All five analyses yield satisfactory $p$-values larger than our a-priori threshold
of $3\%$. The $\chi^2$ and $p$-values are collected in \autoref{tab:results:gof}, alongside the evidence
$\log(Z)$ and our results for the $\psi(3770)$ mass and width.
The best-fit points for all analyses pass the viability tests discussed in \autoref{sec:setup:analysis}.\\

Our minimal fit model provides an excellent description of the data, with a $p$-value of $46\%$ percent.
To study model uncertainties for our fit parameters and derived quantities, we continue to investigate the remaining fit models.
We first compare the three models that use two distinct effective channels, i.e., the models ``no background'', ``background'' and ``$q_0$ variation''.
Although the ``no background'' model features the same $\chi^2$ value as the ``minimal'' model, it is strongly disfavoured with respect to the latter according to Jeffreys' interpretation of the Bayes factor of $B(\text{no background},\allowbreak \text{minimal}) \simeq 1/20$.
The ``background'' model substantially improves the quality
of the fit by decreasing $\chi^2$ by $13$ at the expense
of $3$ additional parameters. This leads to a preference
in terms of the likelihood-ratio test by about $3\sigma$
while being as efficient in the description of the data
as the ``minimal'' model with a Bayes factor of $\sim 0.7$.
This is contrast to the ``$q_0$ variation'' model, which
sees a similar improvement to the $\chi^2$ value;
however, it is disfavoured decisively by a Bayes factor
of $3\cdot 10^{-6}$ with respect to the ``minimal'' fit
model.
We therefore consider the results obtained in the ``background'' model as our nominal results in the case of two distinct
effective channels.

The model ``interference'' with its description of the data with a single, interfering effective channel
gives an equivalent fit quality compared to the ``background'' model but it is slightly less efficient
in its description of the data: the Bayes factor yields
$$
    B(\text{background, interference}) \simeq 3.3\,,
$$
which is ``barely worth mentioning''
according to Jeffreys' interpretation of the Bayes factor.\\

We thus see that the ``background'' and the ``interference'' models both provide an excellent description of the data although the former is somewhat favoured.
The distinction of the two is that in the``background'' model the two vector resonances included in the model cannot interfere via the \nonDDbar channels while in the ``interference'' model they can.
In this sense, the two models provide two extreme scenarios: one assumes that the decay channels of the resonances are all distinct, the other that they are identical.
We therefore expect the spread of our results in either model to cover the true physical results.
A further investigation of this issue would mandate a fit to the respective set of physical exclusive \nonDDbar modes.

The posterior samples for both models are available in form of machine-readable files upon request.
No sizable departure from Gaussian distributions are found in the posterior and all samples pass the viability tests discussed in \autoref{sec:setup:analysis}.

\begin{figure*}[t]
    \centering
    \includegraphics[width=0.48\textwidth]{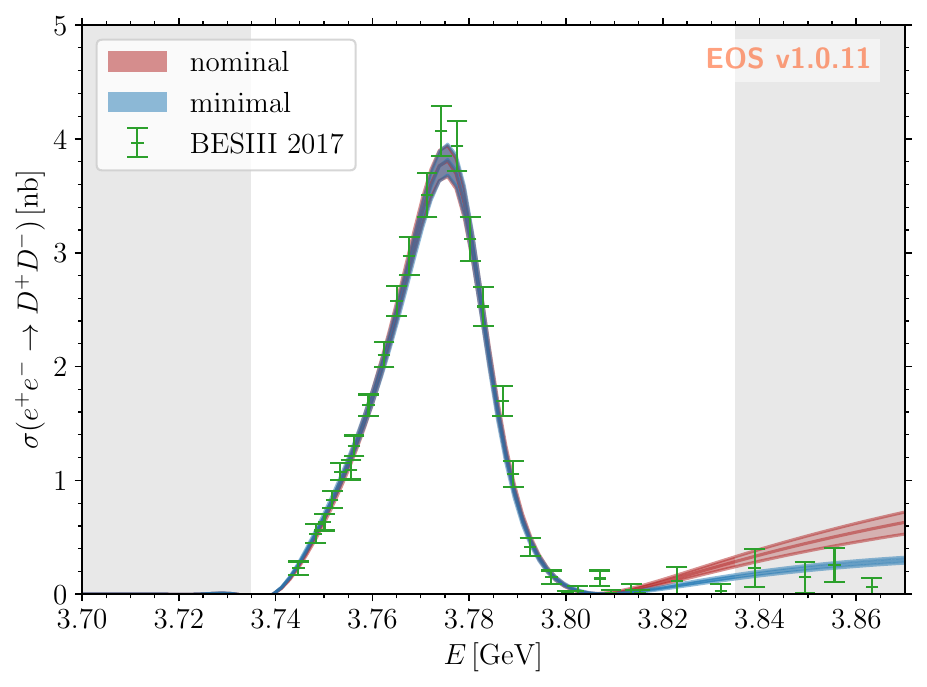}
    \includegraphics[width=0.48\textwidth]{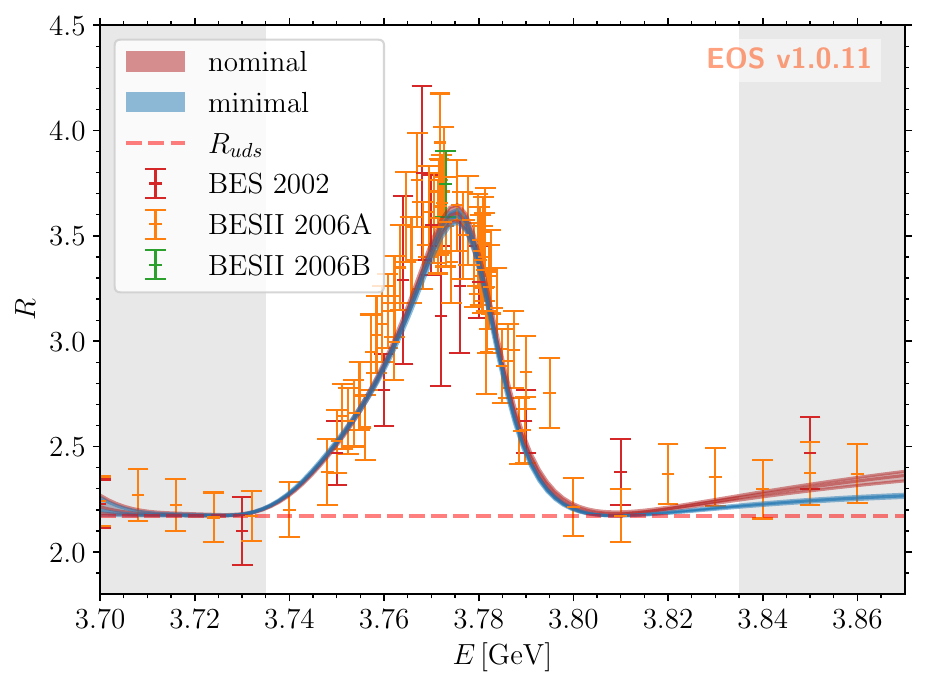}
    \includegraphics[width=0.48\textwidth]{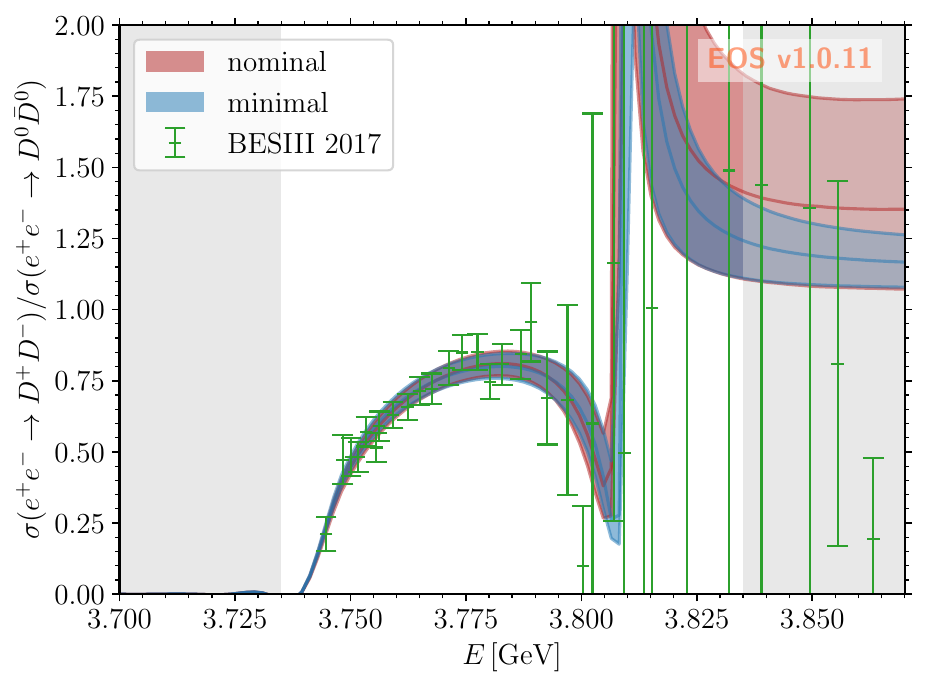}
    \includegraphics[width=0.48\textwidth]{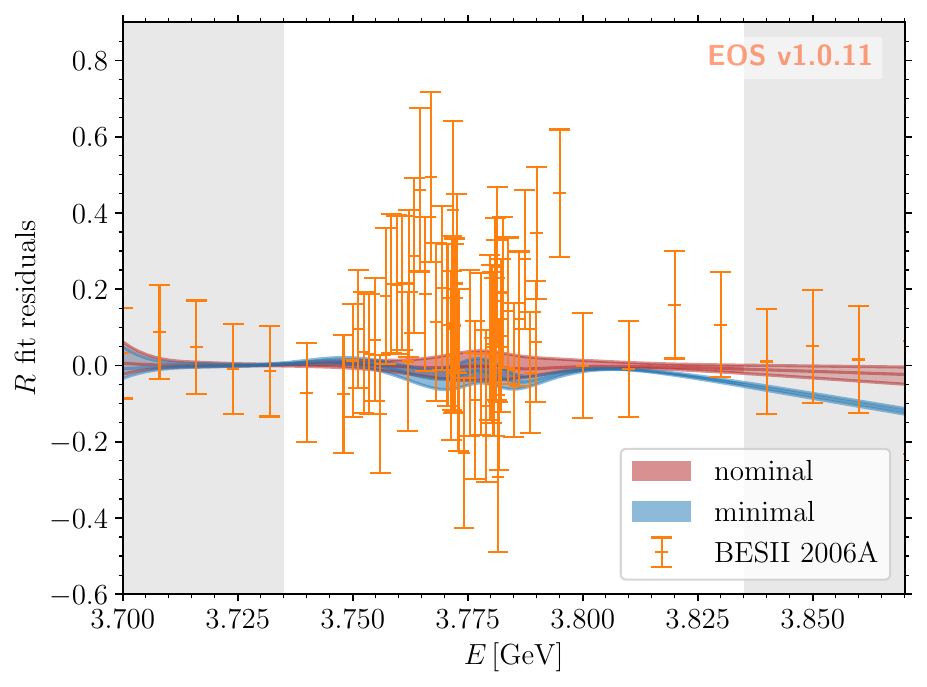}
    \caption{\label{fig:result:NominalFit}
        Predictions of our minimal and nominal models (the ``background'' and ``interference'' models give indistinguishable shapes) for a couple of
        observables in the region of the $\psi(3770)$ resonance,
        compared to the experimental measurement performed by the BES experiment.
        Shaded areas are not used in the fit.\\
        \emph{Top left:} Cross-section of the $e^+e^- \to D^+D^-$ scattering.
        \emph{Top right:} $R$-ratio.
        \emph{Bottom left:} Ratio of the cross-sections of $e^+e^- \to D^+D^-$ and $e^+e^- \to D^0\bar{D}^0$.
        The experimental points are given for illustrative purpose and
        neglect the experimental correlations between the $D^+D^-$ and $D^0\bar{D}^0$ final states.
        \emph{Bottom right:} Residuals of the fit of the $R$-ratio, the nominal ``background'' model is used for the subtraction
        and compared with the minimal model and experimental data.
    }
\end{figure*}

We present the predictions of both models in \autoref{fig:result:NominalFit}.
In the upper plots, the cross-section of $e^+e^- \to D^+D^-$ scattering and the $R$-ratio are compared to the experimental data used in the fit.
The shaded regions indicate the data not used in the fit.
In the bottom right plot, we show the fit residuals for the $R$-ratio.
It is obtained by subtracting the $R$-ratio line shape of our nominal best fit from both the experimental data and the predictions in the ``background'' and ``interference''.
The residual excess of the data around $E = 3.765 \GeV$ motivated the interpretation of the $\psi(3770)$ as a double pole~\cite{Ablikim:2008zzc}.
Our results show that the data can be fully explained by interference effects between the $\psi(3770)$ with the $\psi(2S)$ resonance, an effect not included in Ref.~\cite{Ablikim:2008zzc}.
\footnote{%
    It is not clear to us if the analysis presented in Ref.~\cite{Ablikim:2008zzc} uses further experimental
    data that is not publicly available.
}

Our results deviate from those of Ref.~\cite{Coito:2017ppc} in various aspects: while in our
case the parameter in the regulator functions does not play a significant role as is expected,
since the line shape should be dominated by the resonance itself, in that
work it was determined with a $1\%$ accuracy. This means that in Ref.~\cite{Coito:2017ppc} the regulator
plays a crucial role to shape the resonance. Our fits only need the well established
$\psi(2S)$ and $\psi(3770)$ as poles of the amplitude, while the fits of Ref.~\cite{Coito:2017ppc},
where the $\psi(2S)$ was omitted, dynamically generate an additional pole. The authors
stress that this emergence is unavoidable, if one wants to get a 
good description of the data.
However, our analysis shows that high-accuracy descriptions of the data are possible even in scenarios without that additional pole, as long as the
$\psi(2S)$ is included in the analysis.
Thus, we may conclude that the interplay of an additional pole
with that of the $\psi(3770)$ is indeed necessary to understand the line shape of the latter, however,
this additional pole can well be an established charmonium state.

\paragraph{Mass and width}~Within both of our nominal fit models, we obtain for the physical mass and total decay width
of the $\psi(3770)$ identical results:
\begin{equation}
\begin{aligned}
    M_{\psi(3770)}
        & = 3778.8 \pm 0.3 \MeV \\
    \Gamma_{\psi(3770)}
        & = 25.0 \pm 0.5 \MeV\,.
\end{aligned}
\end{equation}
These values are consistent with those
extracted 
in Ref.~\cite{Coito:2017ppc} 
\begin{equation}
\begin{aligned}
    M_{\psi(3770)}^{\text{\cite{Coito:2017ppc}}}
        & = 3777.0 \pm 1.0 \MeV \\
    \Gamma_{\psi(3770)}^{\text{\cite{Coito:2017ppc}}}
        & = 24.6 \pm 1.0 \MeV\,.
\end{aligned}
\end{equation}
The stability of the pole location is very reassuring, given that there are significant differences in 
the actual modelling of the non-$\psi(3770)$ physics between our work and Ref.~\cite{Coito:2017ppc},
as outlined above.

We remind that our results are obtained from a \Kmatrix analysis. They are therefore not expected to reflect
the parameters extracted from Breit-Wigner analyses, such as the one of Ref.~\cite{Shamov:2016mxe} or the
world average quoted in the PDG review~\cite{Workman:2022ynf}. Nevertheless, we provide these respective results
here for convenience
\begin{equation}
\begin{aligned}
    M_{\psi(3770)}^{\text{\cite{Shamov:2016mxe}}}
        & = 3779.8 \pm 0.6 \MeV \\
    \Gamma_{\psi(3770)}^{\text{\cite{Shamov:2016mxe}}}
        & = 25.8 \pm 1.3 \MeV\,,\\
\end{aligned}
\end{equation}
and
\begin{equation}
\begin{aligned}
    M_{\psi(3770)}^{\text{\cite{Workman:2022ynf}}}
        & = 3778.1 \pm 0.7 \MeV \\
    \Gamma_{\psi(3770)}^{\text{\cite{Workman:2022ynf}}}
        & = 27.5 \pm 0.9 \MeV\,.
\end{aligned}
\end{equation}
We find both the mass and the total decay width to be quite compatible with the literature.
Given the variety of theoretical approaches to describe the data, we do not consider it meaningful
to quote a statistical significance for the deviations.

\begin{figure*}[t]
    \centering
    \includegraphics[width=.6\linewidth]{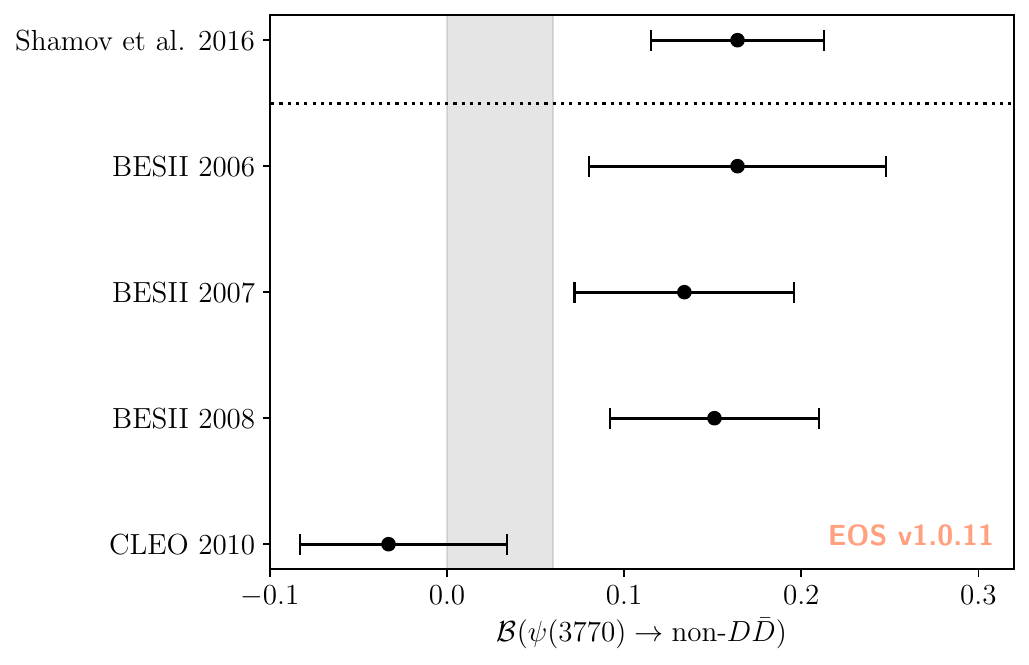}
    \caption{ \label{fig:results:br_to_nonDD}
        Comparison of our upper bound on the branching ratio of the $\psi(3770)$ to \nonDDbar final states at $90\%$ probability (grey band)
        with results from the phenomenological literature~\cite{Shamov:2016mxe} above the dotted line
        and the experimental literature~\cite{BES:2008vad,Ablikim:2007zz,BES:2006fpf,CLEO:2005mpm}
        below.
        Although they use different analysis techniques, the three results by the BESII experiment are not statistically independent.
    }
\end{figure*}

\paragraph{Branching ratio of the $\psi(3770)$ to \nonDDbar}~As already discussed in the
literature~\cite{Shamov:2016mxe,BES:2008vad,Ablikim:2007zz,BES:2006fpf,CLEO:2005mpm},
the combined analysis of inclusive and exclusive measurements
allows for a non-vanishing coupling of the $\psi(3770)$ to \nonDDbar channels, i.e., yielding 
$\mathcal{B}(\psi(3770) \to \nonDDbar)$ at the level of a few percent. Our results for this branching ratio are
presented in the last row of \autoref{tab:results:gof}.
Finding good agreement between the upper bounds in
our two nominal fit models, we summarize our finding as
\begin{align}
    \mathcal{B}(\psi(3770){\to} \nonDDbar)
        &< 6\% \text{ at 90\% probability}\,.
\end{align}
We juxtapose this results with those quoted in the literature in \autoref{fig:results:br_to_nonDD}.
We find that our result is systematically lower than
what is found in the literature, with the exception
of the results of Ref.~\cite{CLEO:2005mpm}.
We mention in passing that using the exclusive observed cross section instead of the exclusive Born cross section
leads to an artificially enhanced \nonDDbar contribution
of the level of $\sim 20\%$.

\paragraph{Isospin symmetry at the $\psi(3770)$ pole}~The $\psi(3770)$ resonance
lies just above the $D^0\bar{D}^0$ threshold ($\sqrt{s} \sim 3.73\GeV$) and the $D^+D^-$ threshold ($\sqrt{s} \sim 3.74\GeV$).
It is therefore sensitive to the differences in phase space volume between the two channels,
leading to an apparent violation of isospin symmetry in the ratio of the exclusive cross sections;
see the bottom left plot of \autoref{fig:result:NominalFit} for an illustration.
We prefer to probe the degree of isospin symmetry violation at hand of a quantity that is unaffected
by these phase space effects.
To this end, we investigate the ratio of the bare couplings between this resonance and either of the two channels.
Unbroken isospin symmetry would yield unity, with symmetry breaking corrections being naturally suppressed by
powers of $\alpha_e$ and $(m_u - m_d) / m_s$.

We find the ratio of bare couplings to be
\begin{equation}
    g_{D^0\bar{D}^0}^{\psi(3770)} / g_{D^+D^-}^{\psi(3770)} = 0.99 \pm 0.03 \,,
\end{equation}
showing no sign of isospin symmetry violation in these decays in either model.
We therefore conclude that the structure shown in the bottom plot of \autoref{fig:result:NominalFit} originates
from the difference in the phase space volumes.
Our finding is in tension with findings in the literature~\cite{Ishikawa:2023bnx,Julin:2017jcl}, which are obtained
by fitting a Breit-Wigner-like line shape to the $\psi(3770)$ spectrum,
but in line with the findings of Ref.~\cite{Coito:2017ppc}. In addition, we determine
the isospin ratio of the bare $D\bar{D}$ couplings to the $\psi(2S)$ resonance to be $1.02 \pm 0.10$,
which is also compatible with unity with substantially larger uncertainties.
The larger uncertainty obtained in this ratio is likely due to the fact that we are not fully modelling the $\psi(2S)$ resonance,
as described in \autoref{sec:setup:framework}.

\section{Summary and Outlook}
\label{sec:summary}

In this paper we have performed a coupled-channel analysis of $\eetoopencharm$ processes in a window around the $\psi(3770)$.
Our analysis compares different models based on the \Kmatrix framework.
We find that the now available high-resolution measurements by the BES, BESII, and BESIII experiments
can be described very well within our models.
We have found no indication for a sizable branching ratio to \nonDDbar final
states.
Modelling these \nonDDbar channels with a single effective $P$-wave channel,
we set an upper bound
\begin{equation*}
    \mathcal{B}(\psi(3770) \to \nonDDbar) < 6\% \text{ at 90\% probability}\,.
\end{equation*}
Our result is compatible with but systematically smaller than nearly all other determinations of
this branching fraction.
In recent years, various vector states were identified as good candidates for exotic states beyond the quark model---see Refs.~\cite{Lebed:2016hpi,Esposito:2016noz,Olsen:2017bmm,Guo:2017jvc,Brambilla:2019esw,Chen:2022asf} for recent reviews.
However, given our results we see no reason to question a dominant $\bar cc$ nature of the $\psi(3770)$.
Note that hadronic loops that drive e.g. the emergence of hadronic molecules~\cite{Guo:2017jvc}
are suppressed near threshold since they appear in a $P$-wave.

In the course of our analysis, we have struggled at times with the lack of statistical constraints
on the electron couplings $g_{e^+e^-}^r$\,. For this coupling to the $\psi(2S)$ we had to resort
to external determinations of the partial width $\Gamma(\psi(2S)\to e^+e^-)$.
We would like to point out that this caveat could be overcome by using measurements of the
cross section $e^+e^- \to \mu^+\mu^-$ in our phase space of interest, which are currently not available at the level
of precision we require.

We look forward to future work in this field, where we plan to extend our analysis to
larger values of $\sqrt{s}$ and, accordingly, to both additional channels and resonances.
This extension will be essential for an envisaged phenomenological application: the transfer
of the line shape information for the vector charmonia from measurements of $e^+e^- \to \text{hadrons}$
cross sections to theoretical predictions of exclusive $b\to s\ell^+\ell^-$ decays.
A sketch of this application is provided in the appendix of this work.
It is presently unclear if this application can be achieved without non-public information on the
experimental measurements, and we hope that this work reinvigorates interest amongst our experimental
colleagues.

\section*{Acknowledgments}

We thank Wolfgang Gradl and Leon Heuser for helpful discussions.
We further thank Chang-Zheng Yuan for pointing out that we used the data from Ref.~\cite{Julin:2017jcl} inconsistently in an earlier
version of this paper.
CH and DvD acknowledge support by the German Research Foundation (DFG)
through the funds provided to the Sino--German Collaborative Research Center TRR110 ``Symmetries and the Emergence of Structure in QCD'' (DFG Project-ID 196253076 -- TRR 110).
The work of SK and DvD was further supported by the DFG within the Emmy Noether Programme under grant DY-130/1-1.
DvD acknowledges ongoing support by the UK Science and Technology Facilities Council
(grant numbers ST/V003941/1 and ST/X003167/1).

\appendix
\renewcommand\appendixname{}

\section{\texorpdfstring{Relations to Non-local Form Factors in $\boldsymbol{b\to s\ell^+\ell^-}$}{Relations to Non-local Form Factors in b -> s l l}}
\label{app:non-local}

Non-local hadronic matrix elements in exclusive $b\to s\ell^+\ell^-$ processes pose a major source
of systematic uncertainty to their theoretical predictions~\cite{Gross:2022hyw}.
They have been the focus of theoretical developments for the past decade~\cite{Khodjamirian:2010vf,Khodjamirian:2012rm,Jager:2012uw,Ciuchini:2015qxb,Bobeth:2017vxj,Gubernari:2020eft,Gubernari:2022hxn}.
Using $\bar{B}\to \bar{K}\ell^+\ell^-$ processes as an example for definiteness,
a common definition of the dominant (charm-induced) non-local\footnote{%
    Here and in the jargon of the rare $b\to s\ell^+\ell^-$ decays, ``non-local'' refers to the fact
    that the operator in \autoref{eq:bsll:def-Hmu} has a non-trivial $x$ dependence,
    opposed to the local $\bar{s}\dots b$ operators whose matrix elements dominate the description
    of theses processes off-resonance.
} contributions reads
\begin{equation}
\begin{aligned}
    \label{eq:bsll:def-Hmu}
    \mathcal{H}^\mu
        & = i\!  \int d^4x\, e^{i q\cdot x} \\
        & \bra{\bar{K}(k)} T\big\{ \bar{c}\gamma^\mu c(x), \sum_{i} C_i O_i(0) \big\} \ket{\bar{B}(q+k)}\,.
\end{aligned}
\end{equation}
Here the $O_{i}$ are a set of local operators in the weak effective theory of mass dimension six and with flavour quantum numbers $sbcc$,
\begin{equation}
    O_i = \big[\bar{s} \Gamma_i b\big]\,\big[\bar{c}\tilde{\Gamma}_i c\big]\,,
\end{equation}
with combined Dirac and colour structures $\Gamma_i$ and $\tilde{\Gamma}_i$;
the $C_i$ are their respective Wilson coefficients.
It is convenient to discuss this hadronic matrix element in terms of its scalar-valued
non-local form factors
\begin{equation}
    \mathcal{H}_{(\lambda)}(q^2)
        = P^\mu(\lambda) \, \mathcal{H}_\mu(q)\,.
\end{equation}
Here $\lambda = 0,\pm 1$ denotes a polarization state of the virtual photon coupling to the
vector current, and $P^\mu(\lambda)$ are suitable projection operators; we refer
to Ref.~\cite{Gubernari:2022hxn} for their definition. We emphasize that the $\mathcal{H}_{(\lambda)}$
are complex-valued functions even below all thresholds in $q^2$. This property emerges
since the $\bar{B}$ meson can decay into an on-shell hadronic state by virtue of the
four-quark operators $O_{i}$; see Ref.~\cite{Khodjamirian:2010vf} for a discussion on this topic.\\

A systematic approach to describing $\mathcal{H}_{(\lambda)}(q^2)$ for $q^2 < 4 M_D^2$
has been developed over the course of the last decade~\cite{Bobeth:2017vxj,Gubernari:2020eft,Gubernari:2022hxn}.
Here, we instead focus on the open-charm region $q^2 \geq 4 M_D^2$.
Common approaches to estimate or describe the non-local
form factors in this region include an operator product expansion (OPE) of the time-ordered product in
\autoref{eq:bsll:def-Hmu}~\cite{Grinstein:2004vb,Beylich:2011aq},
and a Breit-Wigner model of the broad charmonium resonance therein~\cite{Kruger:1996cv,Lyon:2014hpa,Brass:2016efg}.
We propose a different approach based on the $P$-vector formalism that utilizes the
information obtained in the main part of this work.
First, we note that by crossing symmetry the scalar non-local form factors can be related to the scattering
amplitude $\bar{B} K \to e^+e^-$
\begin{equation}
    \mathcal{A}_{\bar{B}K,e^+e^-}
        \sim \sum_{\lambda} L_\mu P_{(\lambda)}^{*\mu} \mathcal{H}_\lambda \,,
\end{equation}
where $L_\mu = \bar{u}_\ell \gamma_\mu v_\ell$ denotes the leptonic current.
Similarly, the $P$-wave amplitude for the processes $\bar{B}\to \bar{K}D\bar{D}$ can be related
to $\bar{B} K \to D\bar{D}$ scattering amplitudes $\mathcal{A}_{\bar{B}K,D\bar{D}}$.
Both of these processes are induced only by the weak interaction. As a consequence, their contributions
to the overall width of the various vector charmonium resonances in the unitarization, for example
through the \Kmatrix approach, are negligible. In such cases, the $P$-vector formalism provides
a convenient approach to parametrize both of the amplitudes mentioned above:
\begin{equation}
    \mathcal{A}_{\bar{B}K,a} = n_a \left[1 - \mathcal{K} \Sigma\right]^{-1} P_{\bar{B}K}(s)\,.
\end{equation}
In the above $P_{\bar{B}K}$ represents the source term,
\begin{equation}
    P_{\bar{B}K}(s) = \sum_{r}^{N_R} \frac{\alpha^r g^r_{\bar{B}K}}{m_r^2 - s} + b_{\bar{B}K}\,
\end{equation}
split into a sum of the same resonances accounted for by the \Kmatrix and a background term $b_{\bar{B}K}$.
As before, $m_r$ and $g_r$ represent bare masses and couplings, and the mass parameters should match those
used in the \Kmatrix analysis.
In contrast to the usual $P$-vector formalism, the couplings $g^r_{\bar{B}K}$ and the background term $b_{\bar{B}K}$
are complex-valued quantities. This can be readily understood from the fact that non-local form factors
(and hence the scattering amplitudes) feature non-vanishing imaginary parts below all thresholds, as discussed above.

\bibliographystyle{JHEP}
\bibliography{references.bib}

\end{document}